\documentclass[twocolumn,superscriptaddress,longbibliography,
	aps,preprintnumbers,prb]{revtex4-1}

\usepackage{graphicx}
\usepackage{bm}
\usepackage{color}
\usepackage{epstopdf}
\usepackage{amsmath}
\usepackage{amssymb}
\usepackage{epstopdf}
\usepackage{soul}
\usepackage{float}
\usepackage{lipsum}
\usepackage[urlcolor=blue,colorlinks=true,citecolor=blue,
	linkcolor=blue,pdfstartview={FitH},bookmarks=false]{hyperref}

\graphicspath{{./fig/}}
\setstcolor{red}
\begin{document}

\title{Engineering nodal topological phases in Ising superconductors by magnetic superstructures}

\author{Szczepan G\l{}odzik}
\email[e-mail: ]{szglodzik@kft.umcs.lublin.pl}
\affiliation{Institute of Physics, M.\ Curie-Sk\l{}odowska University, 
20-031 Lublin, Poland}

\author{Teemu Ojanen}
\email[e-mail: ]{teemu.ojanen@tuni.fi}
\affiliation{Computational Physics Laboratory, Tampere University, P.O. Box 692, FI-33014 Tampere, Finland}

\date{\today}

\begin{abstract}
Recently it was discovered that superconductivity in transition metal dichalcogenides (TMDs) is strongly affected by an out-of-plane spin-orbit coupling (SOC). In addition, new techniques of fabricating 2d ferromagnets on van der Waals materials are rapidly emerging. Combining these breakthroughs, we propose a realization of nodal topological superconductivity in TMDs by fabricating nanostructured ferromagnets with an in-plane magnetization on the top surface. The proposed procedure does not require application of external magnetic fields and applies to monolayer and multilayer (bulk) systems. The signatures of the topological phase include Majorana flat bands that can be directly observed by Scanning Tunneling Microscopy (STM) techniques. We concentrate on NbSe$_2$ and argue that the proposed structures demonstrating the nodal topological phase can be realized within existing technology.   
\end{abstract}


\maketitle

\section{Introduction}

According to the modern approach to condensed matter physics, novel states of matter can be realized in designer systems by combining simpler building blocks. This view implies that our access to new phases of matter and emergent quantum particles is ultimately limited only by our imagination and ability to manipulate matter. The designer approach to 1d topological superconductivity\cite{oreg2010,sau2010} has stirred enormous activity, aiming to fabricate Majorana quasiparticles\cite{Mourik1003,das2012} and harness them in applications. While most of the previous work has targeted gapped phases, here we propose fabrication of a 2d nodal phase with flat Majorana edge bands.

Our proposal is based on two recent breakthroughs, the observation of Ising superconductivity in TMDs\cite{Saito2015,Xi2015,Xi2016,Lu2015,Navarro2016,Huang2018,Barrera2018,Sohn2018} and the discovery of novel  techniques to fabricate 2d magnetic structures on van der Waals materials~\cite{babar2018}. Due to the inversion-breaking structure of monolayers, the quasiparticles experience strong valley-dependent out-of-plane SOC. As a result, superconductivity exhibits remarkable robustness in the presence of large magnetic fields inducing a Zeeman splitting far exceeding the Pauli limit. The importance of the SOC in bulk 2H layer structures has been long overlooked probably because the bulk has inversion symmetry. While the stacked monolayers that make up the bulk exhibit staggered SOC that restores inversion symmetry as a whole, the layers are weakly coupled and quasiparticles in individual layers are subject to strong Ising SOC\cite{Bawden2016}. This is particularly interesting since the pioneering work by Kane and Mele \cite{kane} identified an Ising type SOC in graphene as a crucial ingredient of the quantum spin-Hall phase.           

\begin{figure}
\includegraphics[scale=0.45]{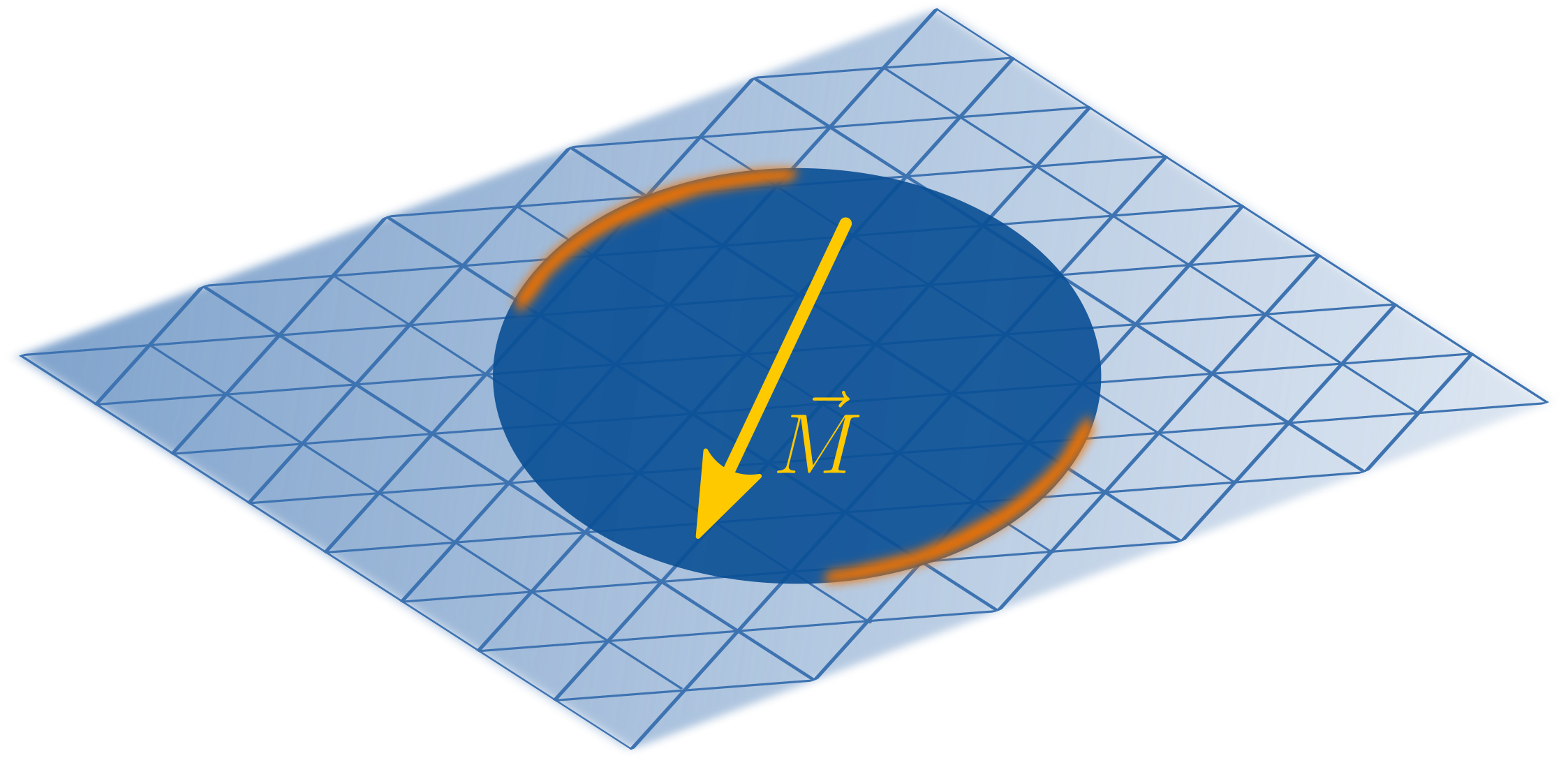}
\caption{Schematic view of the studied system. A magnetic island with in-plane polarization (blue circle) on a top surface of NbSe$_2$ structure. Triangular lattice sites indicate the positions of Nb atoms. The in-plane TMD structure determines the positions of the flat bands. This is independent on the magnetization direction.}
\label{system}
\end{figure}

The rich electronic properties of TMDs have stimulated several proposals for topological and unconventional superconductivity \cite{Zhou2016,Shaffer2019,Yuan2014,Hsu2017,Fischer2018,Taniguchi2012,Oiwa2018,Mockli2018,Sosenko2017,Liu2017,Chen2019}. In Refs.~\onlinecite{He2018,Wang2018} it was proposed that an Ising spin-orbit coupled monolayer TMD (e.g. NbSe$_2$) in the presence of an in-plane magnetic field could realize a topological superconducting state characterized by a nodal bulk and flat Majorana edge bands. In our work we consider how the state can be observed in magnetic structures fabricated on top of an Ising superconductor, treating NbSe$_2$ as a particular realization. We show how magnetic nanostructures on 2H-NbSe$_2$ give rise to the nodal topological phase signalled by the appearance of flat Majorana edge bands. In the light of the ground-breaking success in fabricating 2d magnetic nanostructures on van der Waals materials, our design takes advantage of the latest advances and has several crucial advantages over the magnetic field induced phase~\cite{ohara2018,gong2017,bonilla2018,huang2017}. Our setup does not only remove the necessity of external fields but, importantly, relaxes the requirement of manufacturing monolayer or few layer systems. The proposed topological state engineering works equally well for bulk systems since disruptive orbital effects arising from the in-plane magnetic fields are completely absent in our design. In multilayer systems the topological state is formed in the surface layer in the area in contact with the magnetic structure. The further advantages of the proposed setup include the possibility of fabricating well-defined nanostructures of topological elements. This comes with the benefit that the Majorana flat bands can be directly observed by STM by studying the Local Density of States (LDOS) on magnetic islands. The STM measurements could be employed in resolving the spatial structure of the flat bands, thus providing a smoking-gun signature of the nodal topological phase.  Considering that the process of growing magnetic islands on top of NbSe$_2$ systems is already experimentally on the way, we expect that our predictions will find experimental confirmation in the near future.

\section{Model}

 As a particular model of a TMD we consider a tight-binding representation of 2H-NbSe$_2$. This layered structure consists of stacked units of Nb atoms on a triangular lattice sandwiched by Se layers. To study the appearance and the properties of the nodal topological phase, it is convenient to device a minimum phenomenological model that faithfully displays the essential features of a real material. Since the relevant bands near the Fermi energy mostly derive from Nb $d$ orbitals, the band structure can be qualitatively reproduced by a model with one orbital per site on a triangular lattice. This approximation treats the system effectively as quasi-2d structure, neglecting the less important Se-derived 3d bands near the $\Gamma$ point. While the tight-binding parameters and the band structure vary with the number of layers, the quasi-2d bands remain largely unchanged due to the weak interlayer coupling. Also, experimental observations of magnetic impurities in bulk 2H-NbSe$_2$ highlight the 2d nature of the magnetic subgap spectrum and the qualitative validity of treating the layers as effectively decoupled\cite{Menard2015}.  Thus, while our model is expected to be most faithful in monolayer systems, the experimental evidence suggests it can reasonably capture the qualitative behaviour of the top layer in contact with the magnet in multilayer systems. The Hamiltonian on a triangular lattice with the zigzag edge parallel to the $x$ direction can be written as
$\hat{H}=\hat{H}_{kin}+\hat{H}_{SOC}+\hat{H}_M+\hat{H}_{SC}$, where

\begin{equation}
    \begin{aligned}
    \hat{H}_{kin}&=-t\sum\limits_{\langle \mathbf{i},\mathbf{j}\rangle,\sigma} c_{\mathbf{i}\sigma}^{\dagger}c_{\mathbf{j}\sigma}-\mu\sum\limits_{\mathbf{i},\sigma}c_{\mathbf{i}\sigma}^{\dagger}c_{\mathbf{i}\sigma}-t_2\sum\limits_{\langle\langle \mathbf{i},\mathbf{j}\rangle\rangle,\sigma} c_{\mathbf{i}\sigma}^{\dagger}c_{\mathbf{j}\sigma},\\
    \hat{H}_{SOC}&=-i\lambda\sum\limits_{\langle \mathbf{i},\mathbf{j}\rangle,\sigma,\sigma'} e^{3i\theta_{\mathbf{i}\mathbf{j}}}\sigma_z^{\sigma\sigma'} c_{\mathbf{i}\sigma}^{\dagger}c_{\mathbf{j}\sigma'},\\
    \hat{H}_M&=-\sum\limits_{\mathbf{i},\sigma,\sigma'}[\mathbf{M}(\mathbf{i})\cdot\bm{\sigma}]_{\sigma,\sigma'}c_{\mathbf{i}\sigma}^{\dagger}c_{\mathbf{i}\sigma'},\\
    \hat{H}_{SC}&=\sum\limits_{\mathbf{i}}\Delta(c_{\mathbf{i}\uparrow}^{\dagger}c_{\mathbf{i}\downarrow}^{\dagger}+H.c.).
    \end{aligned}
\end{equation}

Symbols $t$ and $t_2$ correspond to the nearest and next-nearest neighbor hopping and $\mu$ is the chemical potential. Additionally, $\lambda$ parametrizes the out-of-plane Ising-type SOC, the magnetic material gives rise to magnetic splitting $\mathbf{M}$ and $\Delta$ describes the superconducting pairing. Pauli matrices $\sigma$ act on the spin degrees of freedom. Without loss of generality, the magnetization is chosen to point in the $y$ direction $\mathbf{M}=[0,M,0]$. The symbols $\langle \mathbf{i}, \mathbf{j} \rangle$ and $\langle\langle \mathbf{i}, \mathbf{j} \rangle\rangle$ denote the summation over nearest and next-nearest neighbors respectively, $\theta_{\mathbf{i}\mathbf{j}}$ is the angle the vector connecting $\mathbf{i}$ and $\mathbf{j}$ sites makes with the positive $x$ axis (so that $e^{3i\theta_{\mathbf{i}\mathbf{j}}}=\pm1$)and $\bm{\sigma}$ is the vector of the Pauli matrices.

The system is most conveniently analyzed by passing to momentum space and working in the Nambu basis $\psi= (c_{k\uparrow}, c_{k\downarrow}, c^{\dagger}_{k\uparrow}, c^{\dagger}_{k\downarrow})^T$ where the Bogoliubov-de Gennes Hamiltonian becomes 
\begin{equation}\label{BdGH}
 H=E_0(k)\tau_z + E_{SO}(k)\sigma_z + M\sigma_y + \Delta\tau_y \sigma_y.   \end{equation}
The additional set of Pauli matrices $\tau$ act in the particle-hole space. The normal and spin-orbit hopping terms are given by 
\begin{equation}\nonumber
 \begin{aligned}
		E_0&(k) = -2t\bigg[ \cos\big(k_x a) + 2\cos\big(\tfrac{k_x a}{2}\big)\cos\big(\tfrac{k_y a\sqrt{3}}{2}\big)\bigg] \\
			   &- 2t_2\bigg[\cos\big(k_y \sqrt{3}a\big) + 2\cos\big(\tfrac{k_x 3a}{2}\big)\cos\big(\tfrac{k_y \sqrt{3}a}{2}\big) \bigg]-\mu
 \end{aligned}
 \end{equation} 
and 
\begin{equation}
\begin{aligned}
        E_{SO}(k) &= \\
  		2\lambda\bigg[ \sin\big(k_x a\big) &- \sin\big(\tfrac{k_x a}{2} - \tfrac{k_y\sqrt{3}a}{2}\big) - \sin\big(\tfrac{k_x a}{2} + \tfrac{k_y \sqrt{3}a}{2}\big) \bigg].
\end{aligned}
 \end{equation}
 Diagonalization of the Hamiltonian reveals four energy bands:
 \begin{equation} \label{spectrum}
 \begin{aligned}
 E^2=E_0^2 +&E_{SO}^2 +M^2 +\Delta^2 \pm \\
 &2\sqrt{E_0^2(E_{SO}^2+M^2) + M^2\Delta^2}.
 \end{aligned}
 \end{equation}
In the next section we show how the essential normal state features of NbSe$_2$ as well as the nodal topological superconducting state emerge from this model.    

\begin{figure}[t]
\includegraphics[width=\linewidth]{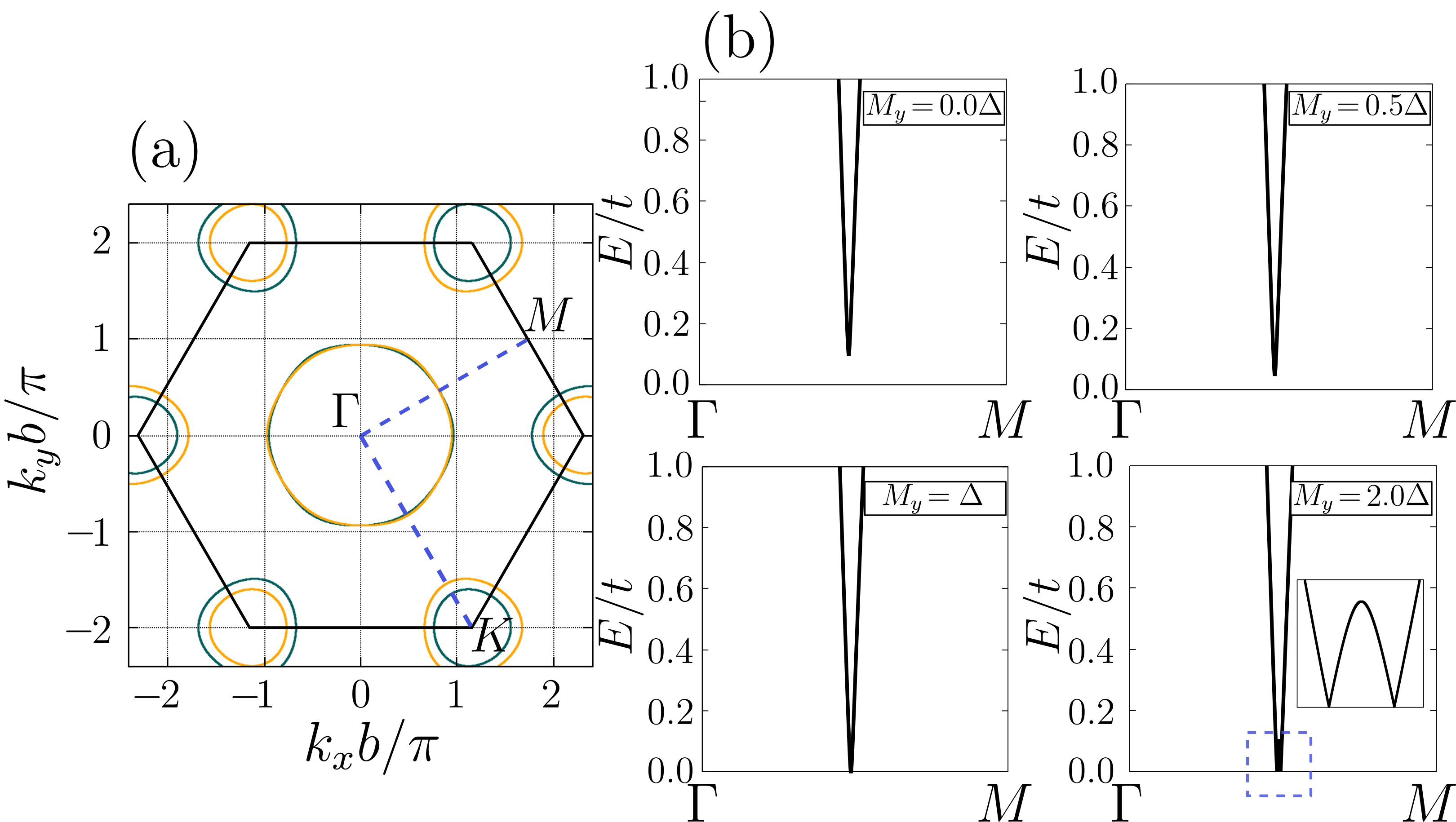}
\caption{Momentum space characteristics of a monolayer TMD. (a) Fermi surface and the Brillouin zone with high symmetry directions as dashed purple lines. Parameters used: $t_2=t$, $\mu=-0.8t$, $\lambda=0.1t$ and $b=\sqrt{3}a$, where $a$ is the lattice constant. (b) The evolution of point nodes along a high symmetry $\Gamma$-$M$ line in the superconducting state with the values of Zeeman field in inset boxes. Same parameters as (a) but with $\Delta=0.1t$. The inset in the last panel shows a close-up of the purple dashed box. }
\label{fsurf}
\end{figure}

\section{Nodal topological superconductivity and flat edge bands}

\subsection{Normal state properties}

 To model the system in the normal state without magnetization, we first set $M=\Delta=0$ and plot the Fermi surface in Fig.~\ref{fsurf}, for a set of parameters that qualitatively reproduces the numerically calculated and experimentally resolved band-structure of NbSe$_2$ \cite{Bawden2016,Wijayaratne2017}. We observe the spin-orbit split $\Gamma$ pocket and more visibly split pockets around $K$ and $K'$ points. Within the second-neighbor hopping model that we use, we are also able to resolve a slight trigonal warping of the pockets. The SOC polarizes the electron spins at $K$ and $K'$ points in the opposite out-of-plane directions. The splitting vanishes along $\Gamma$-$M$ that define mirror symmetric lines. Note that in a multilayer system the spin-orbit coupling is staggered and $\lambda$ has opposite sign in adjacent layers. Nevertheless, the top layer relevant for the topological nodal state in multilayer systems will experience a strong Ising SOC just as a monolayer system. 
 
\subsection{Nodal topological state}

\label{kspace}
 In the presence of superconductivity and magnetization the spectrum is given by Eq.~\eqref{spectrum}. The spectrum is symmetric with respect to zero energy and branches with the minus sign in front of the square root will determine the properties near Fermi energy. For $\Delta>M$ the system is fully gapped while for $M\geq\Delta$ the system becomes gapless with isolated nodal points $k_0$ satisfying $E(k_0)=0$. From the dispersion \eqref{spectrum} we can calculate that at $M=\Delta$ a pair of nodes (with opposite winding) nucleate on the crossing point of the Fermi surface and the $\Gamma$-$M$ lines. This will give rise to six pairs of nodes in the Brillouin zone. By increasing $M$ the nodes of opposite winding move away from the Fermi surface along the $\Gamma$-$M$ lines. This evolution of the nodes is presented in Fig.~\ref{fsurf}b. The nodes cannot be gapped out by small lattice-symmetric perturbations, hinting to a topological nature of the phase. The topological character will be made more explicit below when we discuss the appearance of the flat edge bands. 

\begin{figure}[b]
\includegraphics[scale=0.65]{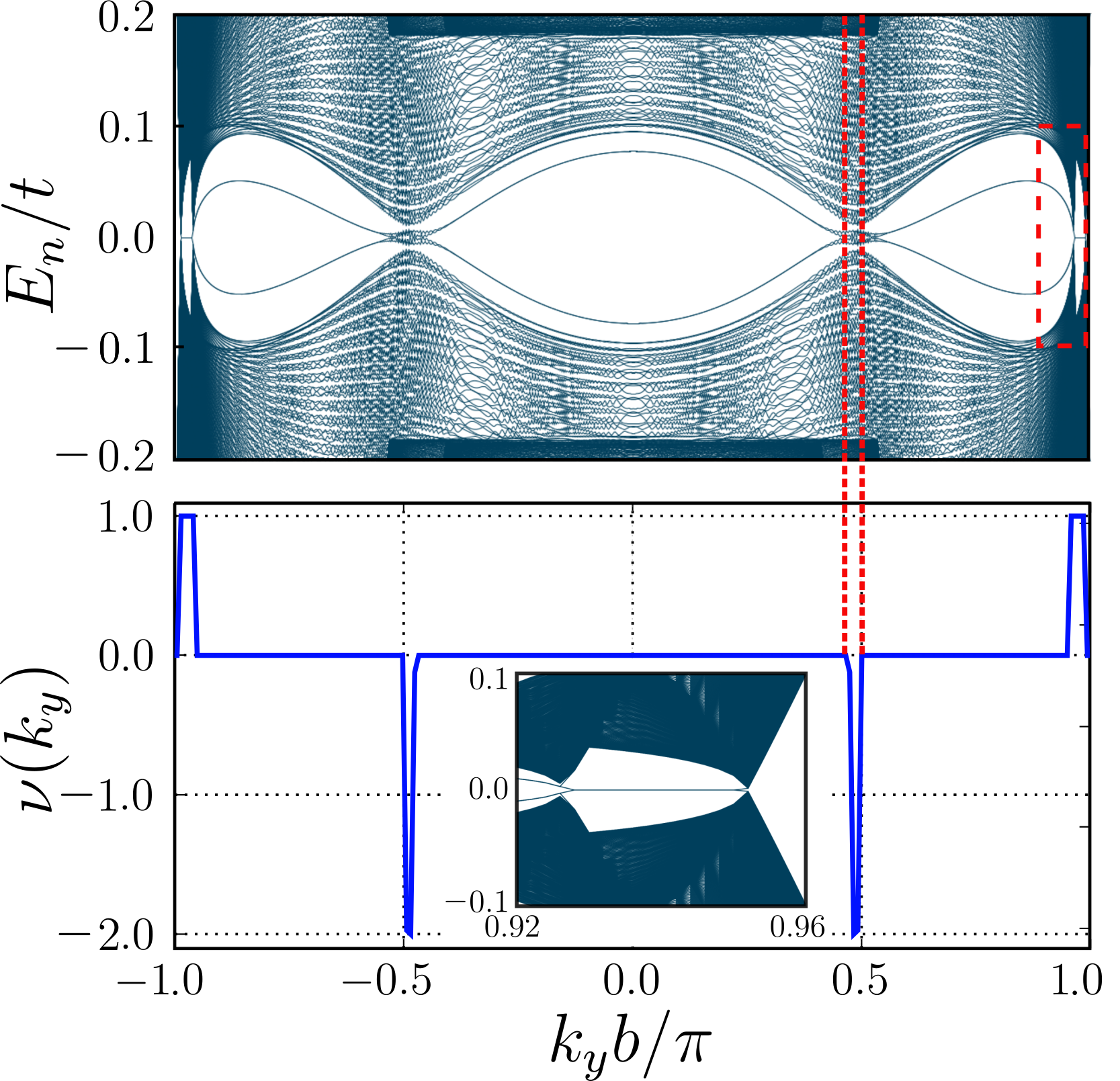}
\caption{Results in the infinite strip geometry.  Top panel: $E(k_y)$ spectrum of the nanoribbon. The number of atoms parallel to the zigzag direction $N=4001$. The red dashed line marks the range of $k_y$ for which we expect a flat band to form. Bottom panel: winding number eq.~\ref{inv}. The inset shows a close-up of the red dashed box. Parameters used: $t_2=t$, $\mu=-0.8t$, $\lambda=0.1t$, $\Delta=0.1t$, $M_y=2\Delta$. }
\label{invrib}
\end{figure}

In analogy to the Fermi arc surface states that connect the surface projections of bulk band-touching points in topological semimetals, the 2d nodal phase supports edge states that connect the edge projections of the nodes. Due to superconductivity, the edge states in the studied system have a Majorana character. The dispersion of the edge bands is flat and the bands appear on edge terminations for which the projections of the nodes with opposite winding do not cancel. Atomic positions on a single layer of NbSe$_2$ form a hexagonal lattice with triangular Nb and Se sublattices. The flat band is the most prominent for an armchair edge termination while it shrinks to a point for a zigzag edge. In order to probe the flat bands, we envision the studied system in an infinite ribbon geometry. We assume periodic boundary conditions (PBC) in $y$ direction, yielding a flat band with maximal extension. So obtained strip spectrum $E(k_y)$ is presented in the top panel of Fig.~\ref{invrib}. The large mismatch of the relevant energy scales $\Delta\sim 0.01t$ would require considering system sizes that are impractical for numerical calculations. Therefore we employ exaggerated parameters $\Delta\sim 0.1t$ for real-space calculations. However, below we argue that the flat bands remain well-defined even for realistic parameters (see Fig.~\ref{nodes}). Close to the nanoribbon BZ edge we see a perfectly flat band at zero energy which connects the nodes. Closer to $k_yb/\pi\simeq \pm 0.5$, there exists another flat band connecting opposite nodes. For the employed parameters, the gap between the bulk states and the edge band is much smaller and the states are less localized to the edge. As a result, the flat band exhibits oscillating departures from the gap center. As shown below, the direct evaluation of the topological invariant protecting the flat band confirms that the inner edge bands exhibit perfect flatness in the large system limit.  

\begin{figure}
     \centering
     \includegraphics[width=0.8\linewidth]{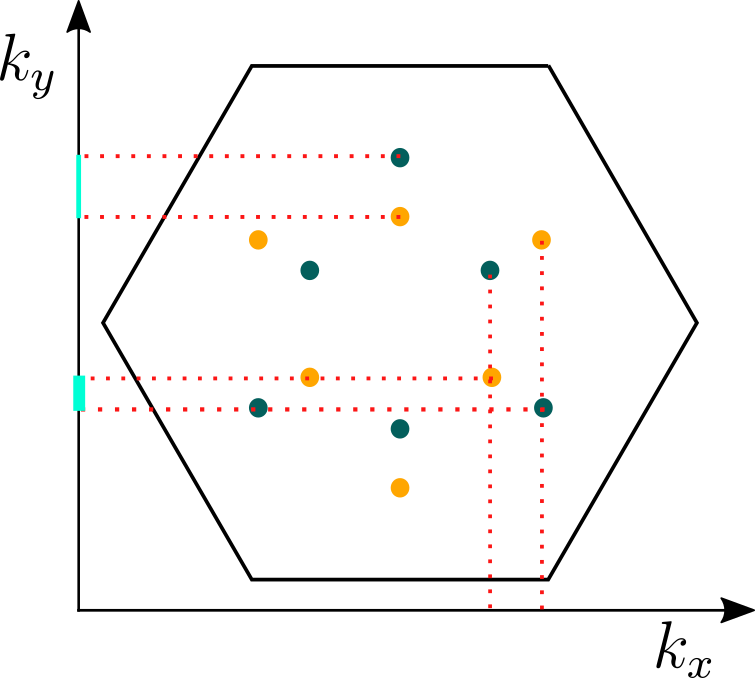}
     \caption{Flat bands are formed between edge projections of nodes with opposite chirality.  The edge parallel to $y$ direction supports four edge bands (two shown) with maximal extension. The band that connects the projections of two pairs of nodes is doubly degenerate. On the edge parallel to $x$ axis the opposite nodes project on top of each other and the flat band vanishes.}
     \label{node}
 \end{figure}

 To elucidate the topological nature of the nodal phase, we calculate the topological invariant protecting the flat bands. This will also provide a definite connection between the edge bands and the edge projection of the nodes. The Hamiltonian \eqref{BdGH} anticommutes with matrix $\mathcal{C}=\tau_y\sigma_x$, hence it belongs to the class BDI, and can be characterized by the winding number in odd spatial dimensions. In a strip geometry we can Fourier transform the 2d tight-binding Hamiltonian  in $y$ direction. Thus, the resulting partially transformed Hamiltonian describes hopping in a 1d chain perpendicular to the strip while $k_y$ is regarded as a parameter. For some intervals of $k_y$ the 1d  Hamiltonian is topologically nontrivial and the 1d chain hosts end states. The flat band can be regarded as the union of the end states of the 1d Hamiltonian. The winding number as a function of $k_y$ can be obtained by evaluating the invariant \cite{volovik}
 
\begin{equation}
 \begin{aligned}
 \nu(k_y) = \frac{1}{2\pi i}\int\limits_{-2\pi}^{2\pi} dk_x Tr \bigg[ \tau_y\sigma_x H^{-1} \partial_{k_x} H \bigg],
 \end{aligned}
 \label{inv}
 \end{equation}
 
 and is illustrated for the studied model in the bottom panel of Fig.~\ref{invrib}. We can see that close to the ribbon BZ edge, where there is a visible flat band, the value of the invariant is $\nu(k_y)=1$. There is also an interval of non-trivial momenta around $k_yb/\pi=0.5$, with higher value of the topological invariant $\nu(k_y) = -2$. The different signs and values of the winding number come from the addition of nodes when projecting them onto the $k_y$ direction as depicted in Fig.~\ref{node}. Physically this means that the flat band is doubly degenerate when $|\nu(k_y)| = 2$. Close to the BZ edge, one pair of nodes adds up and the invariant is equal to unity. The mid-gap bands close to the middle of the BZ that suffer from the proximity of the bulk bands, coincide with invariant value $\nu(k_y) = -2$. Thus they are unambiguously identified as flat bands in the large system limit. 
 
\begin{figure}
\includegraphics[width=\linewidth]{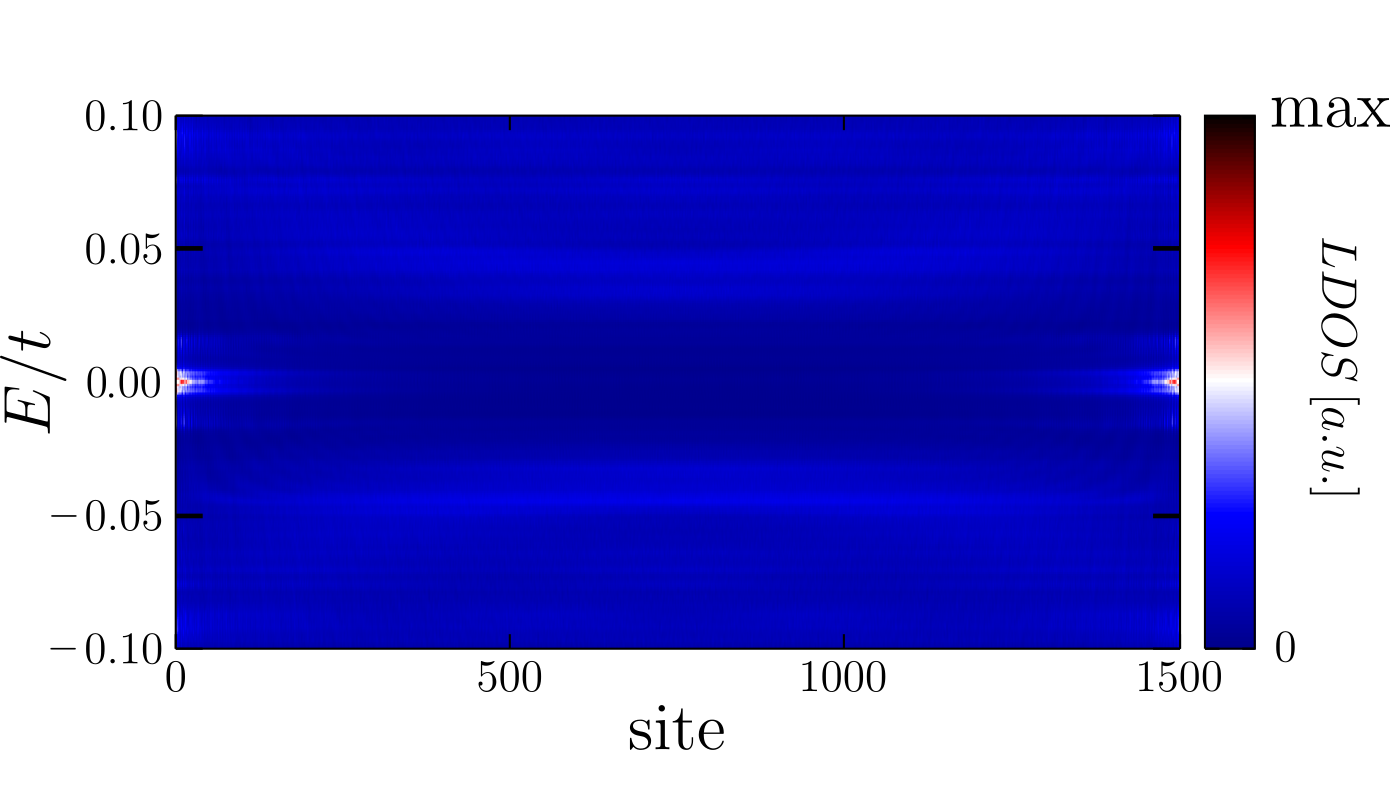}
\caption{The local density of states w.r. to the energy $E$ and the site index in the nanoribbon geometry. Parameters same as in Fig.~\ref{invrib}, expect for $\Delta=0.3t$.}
\label{E_site}
\end{figure}

  To see that the flat bands are indeed localized at the edges of the nanoribbon, we calculated the local density of states (LDOS) as a function of energy and the site index in the direction perpendicular to the PBC
  
  \begin{equation}
  \begin{aligned}
      A(\mathbf{i},E) &= \sum\limits_{k_y,n,\sigma}\bigg[|u_{\mathbf{i}n\sigma}(k_y)|^2\delta\big(E-E_n(k_y)\big)\\
      &+|v_{\mathbf{i}n\sigma}(k_y)|^2\delta\big(E+E_n(k_y)\big) \bigg],
    \end{aligned}
  \end{equation} 
  where we sum over all values of momentum, every state $n$ and both spin directions $\sigma= \uparrow,\downarrow$. The Dirac functions are approximated by Lorentzians with broadening $0.001t$. The result presented in Fig.~\ref{E_site} shows a strong enhancement of the zero-enegy LDOS at the edges of the strip. 

\subsection{Nodal phase on magnetic islands}

\label{rspace}
 The most direct link between our theory and experiments is probably provided by TMD systems with magnetic islands grown on top. These systems, expected to display rich interplay of competing orders, are currently becoming accessible in experiments. The study of coexistence of superconductivity and magnetization has a long history. A single magnetic impurity induces a pair of bound states in the superconducting energy gap. Their shape in real space reflects the symmetry of the underlying lattice, and can extend far away from the impurity depending on the dimensionality of the substrate and other factors like e.g. spin-orbit interaction \cite{Menard2015,Ptok2017}. In the case of a collection of impurities, a Shiba band forms which can undergo a topological phase transition in the deep-dilute limit. Such two dimensional islands of magnetic impurities are recently receiving much attention in extensive theoretical and experimental studies~\cite{Rontynen2015,rontynen2016,Morales2019,Menard2017}, with proposals of engineering hybrids with different dimensionality~\cite{Kobialka2019}. In this context, the present work turns a new leaf on topological state engineering through magnetization, generalizing the previous efforts to obtain gapless topological phases.   

\begin{figure}[t]
\includegraphics[width=\linewidth]{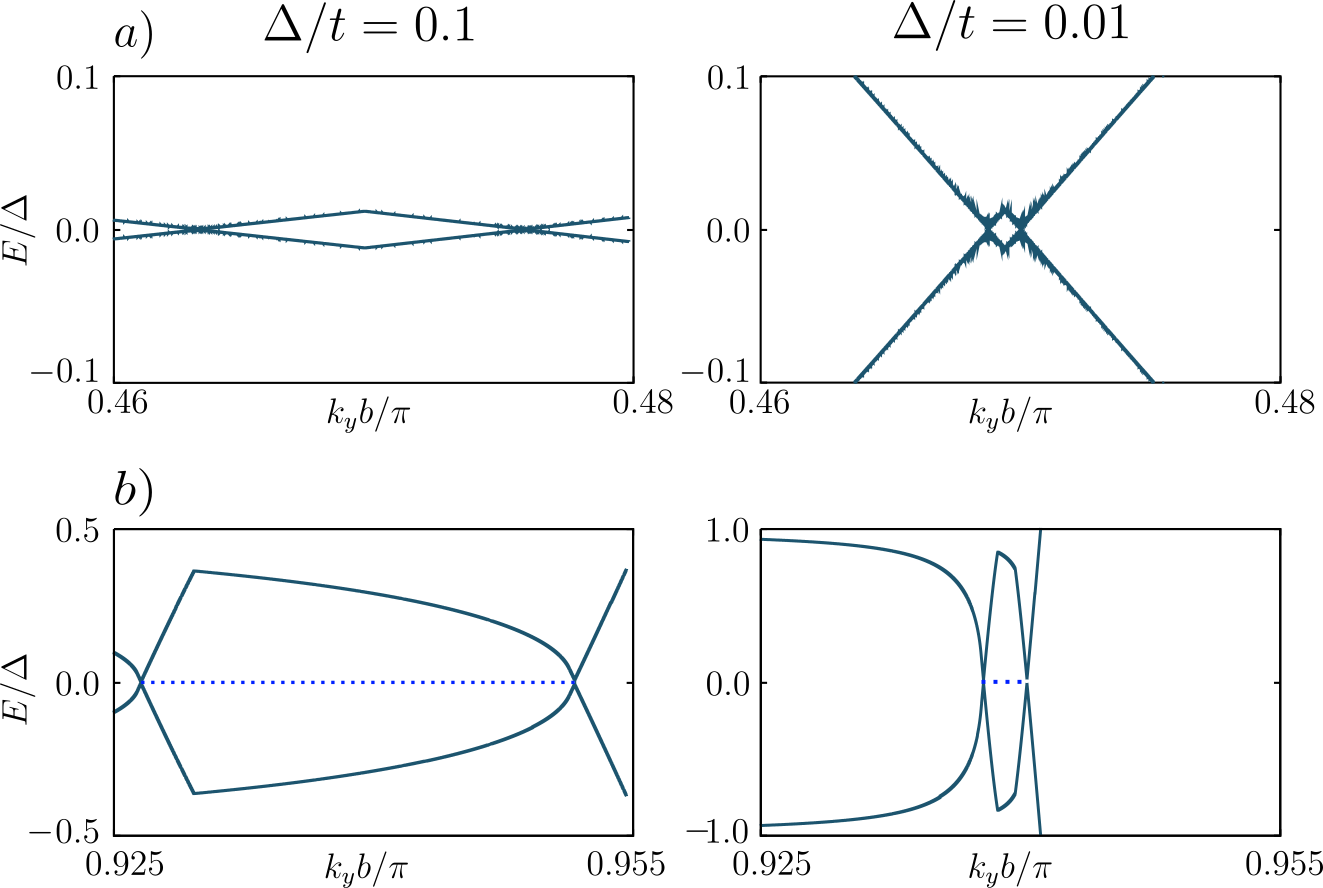}
\caption{Evolution of the point nodes (crossing points) and low-lying spectrum for different values of $\Delta/t$. a) 'Inner' nodes  b) 'outer', close to the Brillouin zone edge nodes. Dotted lines in b) denote the position of flat bands in a finite system. While the length of the flat band is diminished, the gap to the bulk states remains significant.}
\label{nodes}
\end{figure}

\begin{figure}[t]
\includegraphics[width=\linewidth]{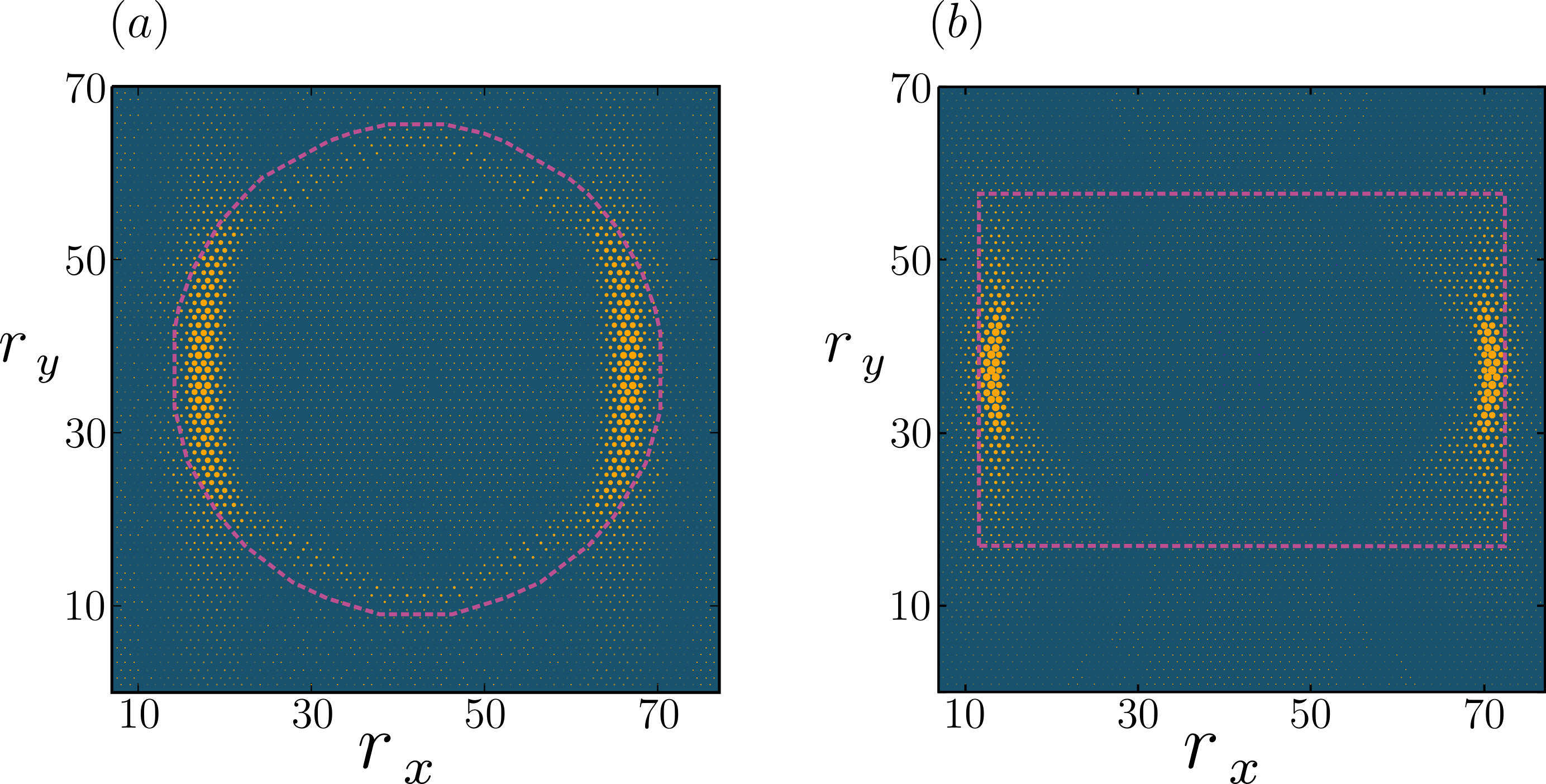}
\caption{Local density of states integrated over the two lowest-lying states, expressed as the size of the yellow dots. Dashed lines outline the shapes of magnetic islands. Parameters used: $t_2=t$, $\mu=-0.8$, $\lambda=0.1t$, $\Delta=0.3t$, $M_y=2\Delta$ }
\label{edgestates}
\end{figure}

For real space calculations we have constructed a triangular lattice consisting of $85$x$86$ atomic sites and applied PBCs in both directions. Then we have applied Zeeman field pointing in the $y$ direction on a collection of sites that comprise the magnetic island. Our approach of fully diagonalizing the lattice model is laden with high computational cost, hence the limited finite size of the lattice. Due to the finite size limitations, the flat band physics can be properly resolved only for the two lowest-lying (closest to zero energy) states, and for the toy parameters, in which the superconducting energy gap is strongly exaggerated ($\Delta \sim t$). However, this is sufficient to demonstrate qualitatively the effects of the nodal phase. To demonstrate that we show in Fig.~\ref{nodes} how the low-lying spectrum between the nodal points evolve as values of $\Delta/t$ are decreasing. One can see that the distance from the gap center to the bulk bands is approximately the same for different values of $\Delta$, and for the outer nodes its value remains a significant fraction of the superconducting gap. This means that the flat bands are always clearly resolved from the bulk states and that the calculations with exaggerated $\Delta/t$ values do not qualitatively change the results. The summed LDOS of the low-energy states is shown in Fig.~\ref{edgestates}. In panel (a) we assumed a circular shape of the magnetic island with radius $r=28a$ from the center of the lattice, whereas in (b) the island has a rectangular shape of the size $60a$\ x\ $40a$, where $a$ is the lattice constant, assumed to be equal to unity. The number of atomic sites comprising the island region is approximately equal in both geometries. The underlying lattice is periodic in every direction, hence the edge of the island is the only edge in our system. The nodal topological phase emerges only when there is an in-plane magnetization, so the edge of the island is the boundary between a gapped trivial phase and the nodal topological phase. Because of the structure of point nodes (as explained in Sec.~\ref{kspace}) and their mutual annulment in the $k_x$ projection, the edge states will never appear in parallel to $x$ direction. We observe that indeed, regardless of the shape of the magnetic island, localized Majorana modes appear only on the edges parallel to $y$ direction. As stressed above, a quantitative study of the edge bands in real space comes with substantial computational cost and is beyond the scope of the present work.   

\subsection{Note on the Rashba effect}

In general, monolayer systems on a substrate and surfaces of a bulk system break the mirror symmetry perpendicular to layers. Therefore a Rashba-type spin-orbit coupling may be nonzero. The Rashba effect, giving rise to an in-plane spin-orbit field, has adversarial effect on the nodal state as it generally gaps out the nodes. The Rashba contribution to Eq.~\eqref{BdGH} can be implemented through the nearest-neighbor hopping term 

\begin{equation}
    \begin{aligned}
    H_R &= E_{R_x}(k)\sigma_x + E_{R_y}(k)\tau_z\sigma_y,\\
        \end{aligned}
\end{equation}    
with
\begin{equation}\nonumber
    \begin{aligned}       
        E_{R_x}(k) &= \alpha_R \sqrt{3}\sin(\tfrac{k_ya\sqrt{3}}{2})\cos(\tfrac{k_xa}{2}),\\
        E_{R_y}(k) & = -\alpha_R\big(\sin(k_xa)+\sin(\tfrac{k_xa}{2})\cos(\tfrac{k_ya\sqrt{3}}{2}) \big),
    \end{aligned}
\end{equation}

where $\alpha_R$ is the Rashba constant. As can be easily verified, the chiral symmetry protecting the edge band is broken and a strict topological protection is not realized. Approximate nature of chiral symmetry, which typically relies on specific directions of microscopic fields, is rather typical. However, the magnitude of the chiral-symmetry breaking is important in assessing how detrimental it is for the observability of the edge modes. A small symmetry-breaking perturbation pushes the edge modes away from the zero energy and they acquire a weak dispersion. Still, sufficiently small gap and weakly broken chiral symmetry cannot destroy the edge bands. In contrast to the Ising SOC which is determined by the lattice structure of TMDs, the Rashba SOC is case specific. Therefore it can also be very weak, especially in multilayer systems where ripples do not play a role.

\section{Discussion and outlook}

The physical realization of the proposed system could be a multilayer or monolayer TMD in contact with magnetic insulating material grown on top. The material should support in-plane magnetization and not perturb the system significantly. A 2d magnetic insulator with a high-quality contact to TMD would be ideal for this purpose. In fact, the requirement to be an insulator is inconsequential since proximity effect can make a thin magnet superconducting. The recent breakthroughs in fabricating 2d magnets down to monolayer thickness on van der Waals systems provide a promising avenue for our proposal~\cite{babar2018,ohara2018,gong2017,bonilla2018,huang2017}. An interesting candidate for the ferromagnet is VSe$_2$ which is a TMD itself and can be epitaxially grown on NbSe$_2$ and other systems of interest. Furthermore, it has been observed that structures based on VSe$_2$ layers exhibit an in-plane magnetization on different substrates~\cite{bonilla2018}. Considering the emerging nature and rapid development of the field of 2d magnets, increasing number of material candidates are likely to emerge soon. In practice the magnetic material also induces a non-magnetic potential which could shift the chemical potential of substrate and thus change the Fermi surface. However, the precise filling is not crucial for realization of the nodal phase.    

The nodal phase can be experimentally identified by observing the Majorana flat bands. As discussed above, the flat bands give rise to enhanced zero-energy LDOS on certain edge terminations on the island. An ideal probe to access this information is STM. In principle, the surface LDOS can be directly measured as a function of energy. This would resolve the flat bands in space and energy. As long as the in-plane magnetization is sufficiently strong to drive islands to the nodal phase, the flat band is most pronounced and suppressed in the same spatial directions for all islands irrespective of the direction of their magnetization. Observation of an enhanced zero energy LDOS on edges with common tangent for multiple islands would constitute a smoking evidence on the nodal phase.        

In the present work we employed NbSe$_2$ as a candidate material for the topological state engineering. However, for the existence of the nodal phase the crucial features are the (quasi-)2d nature of the system, the Ising SOC within a layer, superconductivity and in-plane magnetization. Thus we expect that other TMDs would also provide promising candidates for the proposed system.

In summary, we proposed transition metal dichalcogenides with ferromagnetic structure on top as promising candidates to realize nodal topological superconductivity and flat Majorana edge bands. The systems could be fabricated and analyzed within existing experimental techniques.

\section{Acknowledgements}

This work was supported by National Science Centre (NCN, Poland) grant 2017/27/N/ST3/01762 (S.G.). We would like to thank Peter Liljeroth for initial stimulus of the project and for the numerous discussions regarding experimental realizations. We also acknowledge fruitful discussions with Tadeusz Doma\'nski.  

\bibliography{reference}

\end{document}